\documentclass[10pt,showpacs,showkeys,twocolumn]{revtex4}
\usepackage{amssymb}
\usepackage{amsmath}
\usepackage{graphicx}
\usepackage{multirow}
\usepackage{gensymb}
\usepackage{float}
\usepackage{MnSymbol}
\usepackage{gensymb}

\begin{document}

\title[Ionization of biological molecules by multicharged 
ions]{Ionization of biological molecules by multicharged ions using 
the stoichiometric model}
\author{A. M. P. Mendez, C. C. Montanari, J. E. Miraglia}
\affiliation{Instituto de Astronom\'{\i}a y F\'{\i}sica del Espacio 
(CONICET--UBA), \\ Buenos Aires, Argentina.}


\begin{abstract}
In the present work, we investigate the ionization of molecules of 
biological interest by the impact of multicharged ions in the 
intermediate to high energy range. We performed full non--perturbative 
distorted--wave calculations (CDW) for thirty--six collisional systems 
composed by six atomic targets: H, C, N, O, F, and S --which are the 
constituents of most of the DNA and biological molecules-- and six 
charged projectiles (antiprotons, H, He, B, C, and O). On account of 
the radiation damage caused by secondary electrons, we inspect the 
energy and angular distributions of the emitted electrons from the 
atomic targets. We examine seventeen molecules: DNA and RNA bases, 
DNA backbone, pyrimidines, tetrahydrofuran (THF), and C$_n$H$_n$ 
compounds. We show that the simple stoichiometric model (SSM), which 
approximates the molecular ionization cross sections as a linear 
combination of the atomic ones, gives reasonably good results for 
complex molecules. We also inspect the extensively used Toburen scaling 
of the total ionization cross sections of molecules with the number of 
weakly bound electrons. Based on the atomic CDW results, we 
propose new active electron numbers, which leads to a better universal 
scaling for all the targets and ions studied here in the intermediate 
to the high energy region. The new scaling describes well the available 
experimental data for proton impact, including small molecules. We 
perform full molecular calculations for five nucleobases and test a 
modified stoichiometric formula based on the Mulliken charge of the 
composite atoms. The difference introduced by the new stoichiometric 
formula is less than 3\%, which indicates the reliability of the SSM to 
deal with this type of molecules. The results of the extensive 
ion--target examination included in the present study allow us to assert 
that the SSM and the CDW--based scaling will be useful tools in this area.
\end{abstract}

\keywords{DNA bases, DNA ionization, stoichiometric model, 
multicharged ions, biological molecules, ionization of molecules}
\pacs{34.50Gb, 34.80Gs, 34.80Dp}

\maketitle

\section{Introduction}

The damage caused by the impact of multicharged heavy projectiles on 
biological targets has become a field of interest due to its recent 
implementation in ion--beam cancer therapy. The effectiveness of the 
radiation depends on the choice of the ions. In particular, theoretical 
and experimental studies with different projectiles have concluded that 
charged carbon ions could be the most suitable ions to be used~\cite{Mohamad2017}. 
Nonetheless, the study of such systems represents a challenge from the 
theoretical point of view. 

The ionization of biological molecules by multicharged ions constitutes 
the primary damage mechanism. The most widely used method to predict
such processes is the first Born approximation. At high energies, this 
perturbative method warrants the $Z^{2}$ laws, where $Z$ is the 
projectile charge. However, the damage is concentrated in the 
vicinities of the Bragg peak --at energies of hundreds of keV/amu--, 
precisely where the Born approximation starts to fail. 
Another theoretical issue arises due to the targets themselves; we are 
dealing with complex molecules, and the description of such targets 
represents a hard task for {\it ab initio} calculations. 

Different approaches have been proposed to deal with the ionization of 
molecular targets within the independent atom model. For example, 
Galassi \textit{et al.} \cite{galassi2000} obtain molecular cross 
sections by combining CDW-EIS atomic ones based on the population of 
the molecular orbitals. More recently, L\"udde \textit{et al.} 
\cite{ludde2016,ludde2018} 
propose a combination of atomic cross sections with geometrical screening
corrections.

The objective of this article is to face with two aspects of the ionization
of biological molecules; first, we perform 
more appropriate calculations on the primary damage mechanism, which can 
replace the Born results. Second, we inspect and test a stoichiometric 
model to describe the ionization of molecular targets.

To overcome the limitations of first order perturbative approximations, 
and since the projectiles are multicharged ions, we resort to the 
Continuum Distorted Wave--Eikonal Initial State 
(CDW)~\cite{galassi2000,fainstein1988,miraglia2008,miraglia2009}, which 
includes higher perturbative corrections. Details on the CDW calculation 
are given in Section~\ref{sec:atoms}.
We start from the premise that the ionization 
process is the mechanism that deposits the most significant amount of 
primary energy. Moreover, the residual electrons from the ionization 
are known to be a source of significant local biological damage~\cite{Denifl2011}. 
The secondary electrons are included in Monte Carlo simulations, 
and hence their behavior requires further investigation.
In Section~\ref{subsec:meanener} and \ref{subsec:meanang}, we calculate
the mean energy and angular distributions of the ejected electrons. 
Surprisingly, we found a substantial dependence on the projectile charge, 
which is unexpected in the first Born approximation. 

In Section~\ref{subsec:stoichiometric}, we deal with the molecular 
structure complexity of the targets by 
implementing the simplest stoichiometric model (SSM): the molecules are 
assumed to be composed of isolated independent atoms, and the total 
cross section by a linear combination of stoichiometric weighted atomic 
calculations.
By implementing the CDW and the SSM, we calculate ionization 
cross section of several molecules of biological interest, including 
DNA and RNA molecules, such as adenine, cytosine, guanine, thymine, uracil, 
tetrahydrofuran~(THF), pyrimidine, and DNA backbone, by the impact 
of antiprotons, H$^{+}$, He$^{2+}$, Be$^{4+}$, C$^{6+}$, and O$^{8+}$. 
In Section~\ref{subsec:scaling}, we
test the Toburen scaling rule~\cite{toburen1975,toburen1976}, which 
states that the ratio between the ionization cross section and the 
number of weakly bound electrons can be arranged in a narrow universal 
band in terms of the projectile velocity. We applied this rule to 
several hydrocarbons and nucleobases and noted that the width of the 
resulting universal band could be significantly reduced if we consider 
the number of active electrons in the collision based on the 
CDW results for the different atoms. The new scaling was then tested 
theoretically and by comparison with experimental data available.

The approach SSM considers the atoms in the molecule as neutral, 
which is not correct. In Section~\ref{subsec:molcalculations}, 
we used the molecular electronic structure code 
{\sc gamess}~\cite{gamess} to calculate the excess or defect of 
electron density on the atoms composing the molecules. Then, we modified 
the SSM to account for the departure from the neutrality of the atoms. 
We find that the modified SSM for the DNA molecules does not introduce 
substantial changes in the cross sections.

\section{Theory: Ionization of Atoms}
\label{sec:atoms}

In the present study, we consider six atoms, $\alpha=$ H, C, N, O, P, 
and S, and six projectiles, antiprotons $\bar{p}$, H$^{+}$, He$^{2+}$, 
Be$^{4+}$, C$^{6+}$, and O$^{8+}$. 
Most of the organic molecules are composed of these atoms. Some 
particular molecules also include halogen atoms such as fluorine and 
bromine; ionization cross sections of these elements have been 
previously published~\cite{miraglia2008}.

The total ionization cross sections of these atoms $\sigma_{\alpha}$
were calculated using the CDW. The initial bound and final continuum 
radial wave functions were obtained by using the {\sc radialf} code, 
developed by Salvat and co--workers~\cite{salvat1995}, and a Hartree--Fock 
potential obtained from the Depurated Inversion Method~\cite{mendez2016,mendez2018}. 
We used a few thousand pivot points to solve the Schr\"{o}dinger 
equation, depending on the number of oscillations of the continuum 
state. The radial integration was performed using the cubic spline 
technique. We expand our final continuum wave function as usual,
\begin{equation}
\psi_{\overrightarrow{k}}^{-}(\overrightarrow{r})=\sum_{l=0}^{l_{\max
}}\sum_{m=-l}^{l}R_{kl}^{-}(r)Y_{l}^{m}(\widehat{r})Y_{l}^{m^{\ast }}
(\widehat{k})\,.
\label{eq:contwave}
\end{equation}
The number of angular momenta considered varied from 8, at 
very low ejected--electron energies, up to $l_{\max}\sim 30$, 
for the highest energies 
considered. The same number of azimuth angles were required to obtain 
the four--fold differential cross section. The calculation performed does 
not display prior--post discrepancies at all. Each atomic total cross 
section was calculated using 35 to 100 momentum transfer values, 28 
fixed electron angles, and around 45 electron energies depending on the 
projectile impact energy. 
Further details of the calculation are given in Ref.~\cite{montanari2017}. 

\begin{figure*}[t!]
\centering
\includegraphics[width=0.8\textwidth]{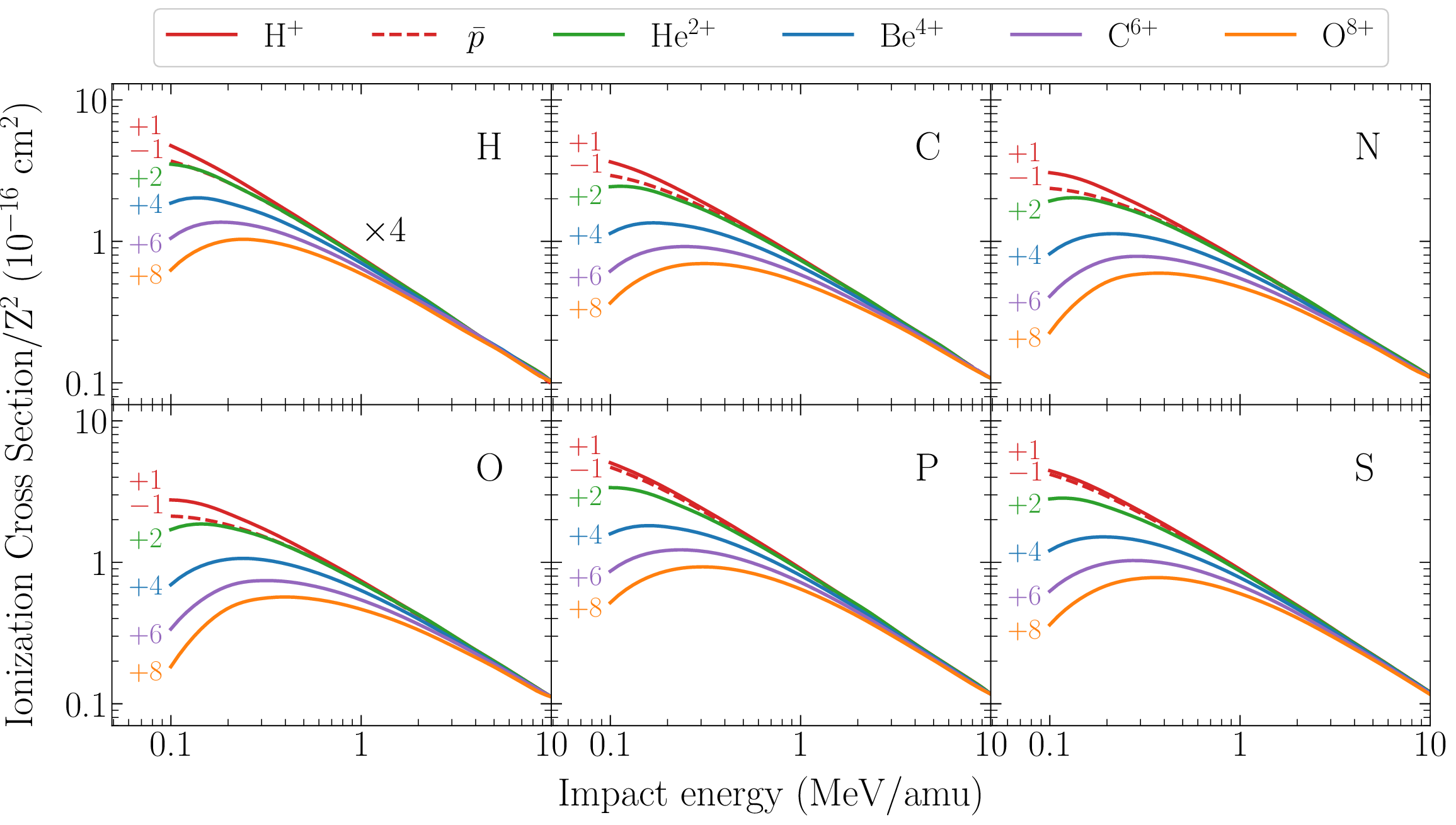}
\caption{Reduced CDW total ionization cross section $\sigma_{\alpha}/Z^2$ 
of six atomic 
targets. The curves are labeled with the charge state corresponding to 
the six multicharged projectiles.}
\label{fig:atomscaling}
\end{figure*} 

We display our total CDW ionization cross sections for the six essential 
elements by the impact of the six projectiles in Fig.~\ref{fig:atomscaling}.
To reduce the resulting 36 magnitudes into a single consistent 
figure, we considered the fact that in the first Born approximation
the ionization cross section scales with the square of the projectile 
charge, $Z^{2}$. The impact energies considered 
range between 0.1 to 10 MeV/amu, where the CDW is supposed 
to hold. In fact, for the highest projectile charges the minimum 
impact energy where the CDW is expected to be valid could be 
higher than 100 keV. We also performed similar calculations with the 
first Born approximation, and we corroborated that it provides quite 
reliable results only for energies higher than a couple of MeV/amu. 
We use the same line color to indicate the projectile charge throughout 
all the figures of this work: dashed--red, solid--red, blue, magenta, 
olive and orange for antiprotons, H$^{+}$, He$^{2+}$, Be$^{4+}$, 
C$^{6+}$, and O$^{8+}$, respectively. Notably, there is no complete 
tabulation of ionization of atoms by the impact of multicharged ions. 
We hope that the ones presented in this article will be of help for 
future works.

Simultaneously, we will be reporting state to state ionization cross 
sections for the 36 ion--target systems considered in the present 
work~\cite{miraglia2019}. A great numerical effort was paid to obtain 
these results, and we expect that they will be useful to estimate 
molecule fragmentation.

\subsection{Emitted electron energies}
\label{subsec:meanener}

In a given biological medium, direct ionization by ion impact accounts 
for just a fraction of the overall damage. Secondary electrons, as well 
as recoil target ions, also contribute substantially to the total damage~\cite{Denifl2011}. 
We can consider the single differential cross section of the shell 
$nl$ of the atom $\alpha$, $d\sigma_{\alpha nl}/dE$, to be a function 
of the ejected electron energy $E$ as a simple distribution 
function~\cite{surdutovic2018}. Then, we can define the mean value 
$\overline{E}_{\alpha}$ as in Ref.~\cite{abril2015},
\begin{eqnarray}
\overline{E}_{\alpha} &=&\frac{\langle E_{\alpha}\rangle}{\langle
1\rangle}=\frac{1}{\sigma_{\alpha}}\sum\limits_{nl}\int dE\,E
\frac{d\sigma_{\alpha,nl}}{dE}\,,  
\label{40} \\
\langle 1\rangle &=&\sigma_{\alpha}=\sum\limits_{nl}\int dE\,
\frac{d\sigma_{\alpha,nl}}{dE}\,,  
\label{50}
\end{eqnarray}
where $\Sigma_{nl}$ takes into account the sum of the different 
sub--shell contributions of the element $\alpha$.

\begin{figure*}[t!]
\centering
\includegraphics[width=0.8\textwidth]{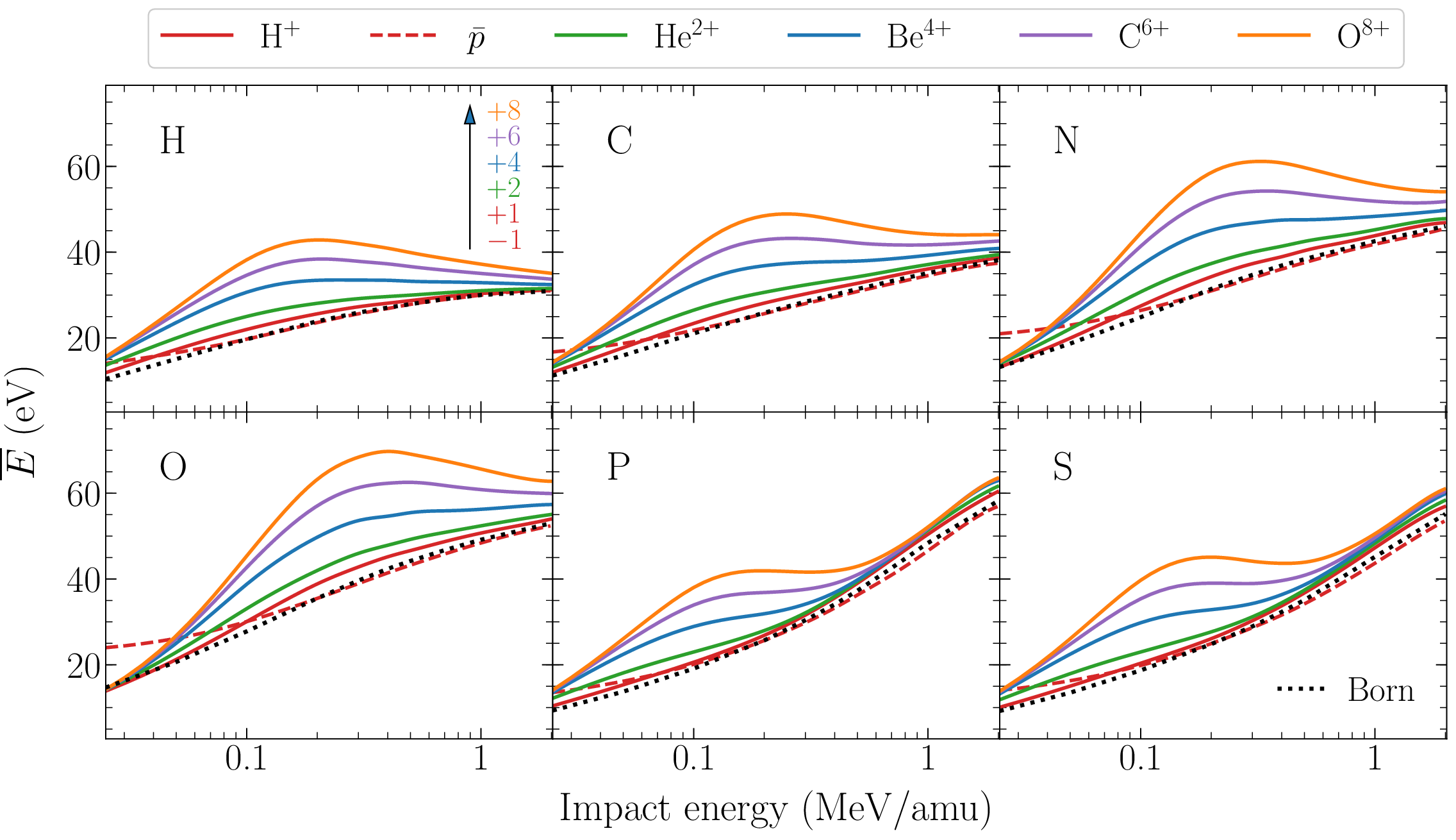}
\caption{Mean emitted energy distribution for ionization by the impact 
of multicharged ions, given by Eq.~(\ref{40}). 
Solid lines for ion charges $+1$, $2+$, $4+$, $6+$ and $8+$, as indicated.
Dashed lines for $\bar{p}$ and dotted line for the Born approximation with $Z=1$.}
\label{fig:emittedener}
\end{figure*} 

The mean emitted electron energies $\overline{E}_{\alpha}$ for H, C, N, 
O, P and S are shown in Fig.~\ref{fig:emittedener}. The range of impact 
velocities was shortened to $v=10$ a.u. due to numerical limitations 
in the spherical harmonics expansion of Eq.~(\ref{eq:contwave}). 
As the impact velocity $v$ increases, so do $\langle E_{\alpha}\rangle$
and $l_{\max}$, which results in the inclusion of very oscillatory 
functions in the integrand. Furthermore, the integrand of
$\langle E_{\alpha}\rangle$ includes the kinetic energy $E$
(see Eq.~(\ref{40})), which cancels the small energy region and 
reinforces the large values, making the result more sensible to large
angular momenta. Regardless, for $v>10$ a.u., the first Born 
approximation holds.

In Fig.~\ref{fig:emittedener}, we estimate $\overline{E}_{\alpha}$ of
the emitted electron in the 10--70 eV energy range,
for all the targets. Our results agree with the 
experimental findings~\cite{surdutovic2018}. As can be noted in the 
figure, the mean energy value is surprisingly sensitive to the 
projectile charge $Z$, which can duplicate the proton results in the 
intermediate region, i.e., 100--400 keV/amu. The effect observed can be 
attributed to the depletion caused by the multicharged ions to the 
yields of low energy electrons. This behavior cannot be found in the 
first Born approximation, where the $Z^2$ law cancels the $Z$ dependence
in Eq.~(\ref{40}). At high energies, $\overline{E}_{\alpha}$ tends to a 
universal value for all ions, as can be seen in Fig.~\ref{fig:emittedener}.

\subsection{Emitted electron angles}
\label{subsec:meanang}

\begin{figure*}[t!]
\centering
\includegraphics[width=0.8\textwidth]{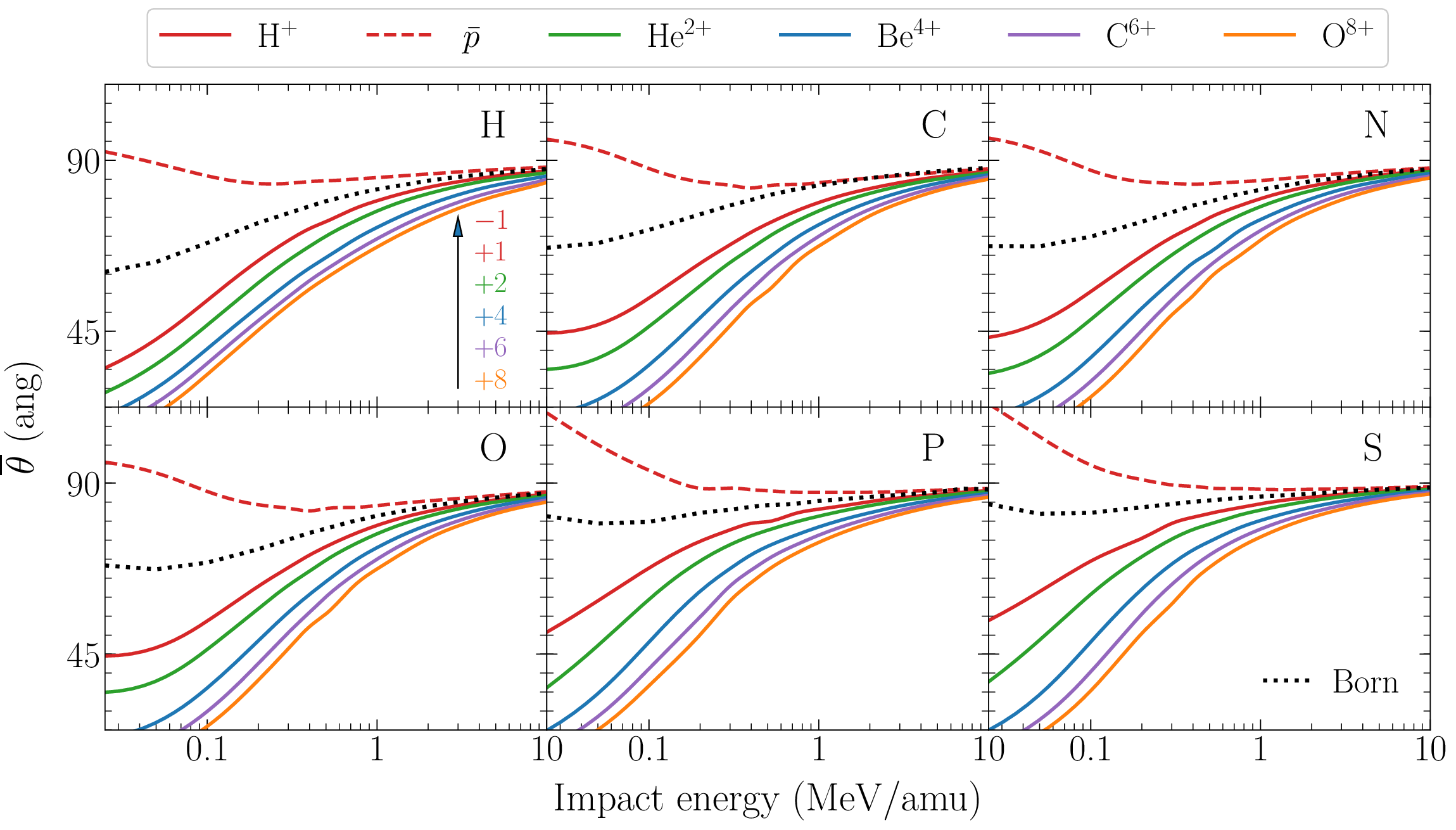}
\caption{Mean emitted angle distribution for ionization by impact of
multicharged ions. Curves as in Fig.~\ref{fig:emittedener}.}
\label{fig:emittedang}
\end{figure*} 

As mentioned before, secondary electrons contribute to the total damage. 
Then, not only the ejection energy is essential but also the angle 
of emission. Once again, we can consider the single differential cross 
section in terms of the ejected electron solid angle $\Omega$, 
$d\sigma_{\alpha,nl}/d\Omega$, to be expressed as a distribution function, 
and the mean angle $\overline{\theta}_{\alpha}$ can be defined as
\begin{eqnarray}
\overline{\theta}_{\alpha}&=&\frac{\langle\theta_{\alpha}\rangle}
{\langle 1\rangle}=\frac{1}{\sigma_{\alpha}}\sum\limits_{nl}
\int d\Omega\,\theta\,\frac{d\sigma_{\alpha,nl}}{d\Omega} \\
\langle 1\rangle &=&\sigma_{\alpha}=\sum\limits_{nl}\int d\Omega\,
\frac{d\sigma_{\alpha,nl}}{d\Omega}\,, 
\end{eqnarray}

The mean emitted electron angles $\overline{\theta}_{\alpha}$ for the 
six atoms and six ions of interest are shown in Fig.~\ref{fig:emittedang}.
A significant dependence of $\overline{\theta}_{\alpha}$ with $Z$ is 
noticed for all the cases. Once again, this effect could not be 
observed in the first Born approximation (dotted line).

For low energy electron emission, the angular dispersion is nearly 
isotropic~\cite{Rudd1992}.
A typical value for the ejection angle considered in the literature is 
$\overline{\theta}_{\alpha}\sim 70\degree$~\cite{surdutovic2018}, and 
it is quite correct in the range of validity of the first Born 
approximation for any target. However, when a distorted wave approximation 
is used, $\overline{\theta}_{\alpha}$ decreases substantially with $Z$ 
in the intermediate energy region, as shown in Fig.~\ref{fig:emittedang}.
The higher the charge $Z$, the smaller 
$\overline{\theta}$ will be. Of course, this effect only holds at 
intermediate energies and not at high impact energies. 

To illustrate this feature, consider the impact of 500~keV C$^{6+}$ on 
oxygen. The first Born approximation predicts emitted electrons with 
mean energies of 46.7 eV and mean angles of 78\degree, while the CDW 
gives 62.5 eV and 60\degree. 
These results imply deeper penetration of the secondary electrons with 
an orientation closer to the direction of the ion. 
We can attribute this forward direction correction to the capture to 
the continuum effect.

Furthermore, Fig.~\ref{fig:emittedang} provides an illustrative 
description of the behavior of antiprotons: the projectile repels the 
electrons, being $\overline{\theta}_{\alpha}\sim 90\degree$. Note the 
opposite effect of proton and antiprotons respect to the first Born 
approximation; this phenomenon constitutes an angular Barkas effect.

\section{Ionization of Molecules}
\label{sec:molecules}
\subsection{The stoichiometric model}
\label{subsec:stoichiometric}

Lets us consider a molecule $M$ composed by $n_{\alpha}$ atoms of the
element $\alpha$, the SSM approaches the total ionization cross section 
of the molecule $\sigma_{M}$ as a sum of ionization cross sections of 
the isolated atoms $\sigma_{\alpha}$ weighted by $n_{\alpha}$, 
\begin{equation}
 \sigma_{M}=\sum\limits_{\alpha}n_{\alpha}\sigma_{\alpha}\,.  
 \label{eq:sumion}
\end{equation}
We classified the molecular targets of our interest in three families: 
CH, CHN, and DNA, as in Table~\ref{tab:families}.

\begin{table}[t]
\begin{center}
\begin{tabular}{|p{0.042\textwidth}|p{0.39\textwidth}|}
\hline
\multirow{2}{*}{CH} 
      & CH$_4$ (methane), C$_2$H$_2$ (acetylene),  \\
      & C$_2$H$_4$ (ethene), C$_2$H$_6$ (ethane), \\
      & C$_6$H$_6$ (benzene) \\
\hline
\multirow{2}{*}{CHN} 
      & C$_5$H$_5$N (pyridine), C$_4$H$_4$N$_2$ (pyrimidine), \\ 
      & C$_2$H$_7$N (dimenthylamine), \\
      & CH$_5$N (monomethylamine) \\
\hline
\multirow{4}{*}{DNA} 
      & C$_5$H$_5$N$_5$ (adenine), C$_4$H$_5$N$_3$O (cytosine), \\
      & C$_5$H$_5$N$_5$O (guanine), C$_5$H$_6$N$_2$O$_2$ (thymine), \\
      & C$_4$H$_4$N$_2$O$_2$ (uracil), C$_4$H$_8$O (THF), \\
      & C$_5$H$_{10}$O$_5$P (DNA backbone), \\
      & C$_{20}$H$_{27}$N$_7$O$_{13}$P$_2$ 
(dry DNA) \\
\hline
\end{tabular}
\caption{Molecular targets studied in this work, classified in three 
families.}
\label{tab:families}
\end{center}
\end{table}

\begin{figure*}[t!]
\centering
\includegraphics[width=0.8\textwidth]{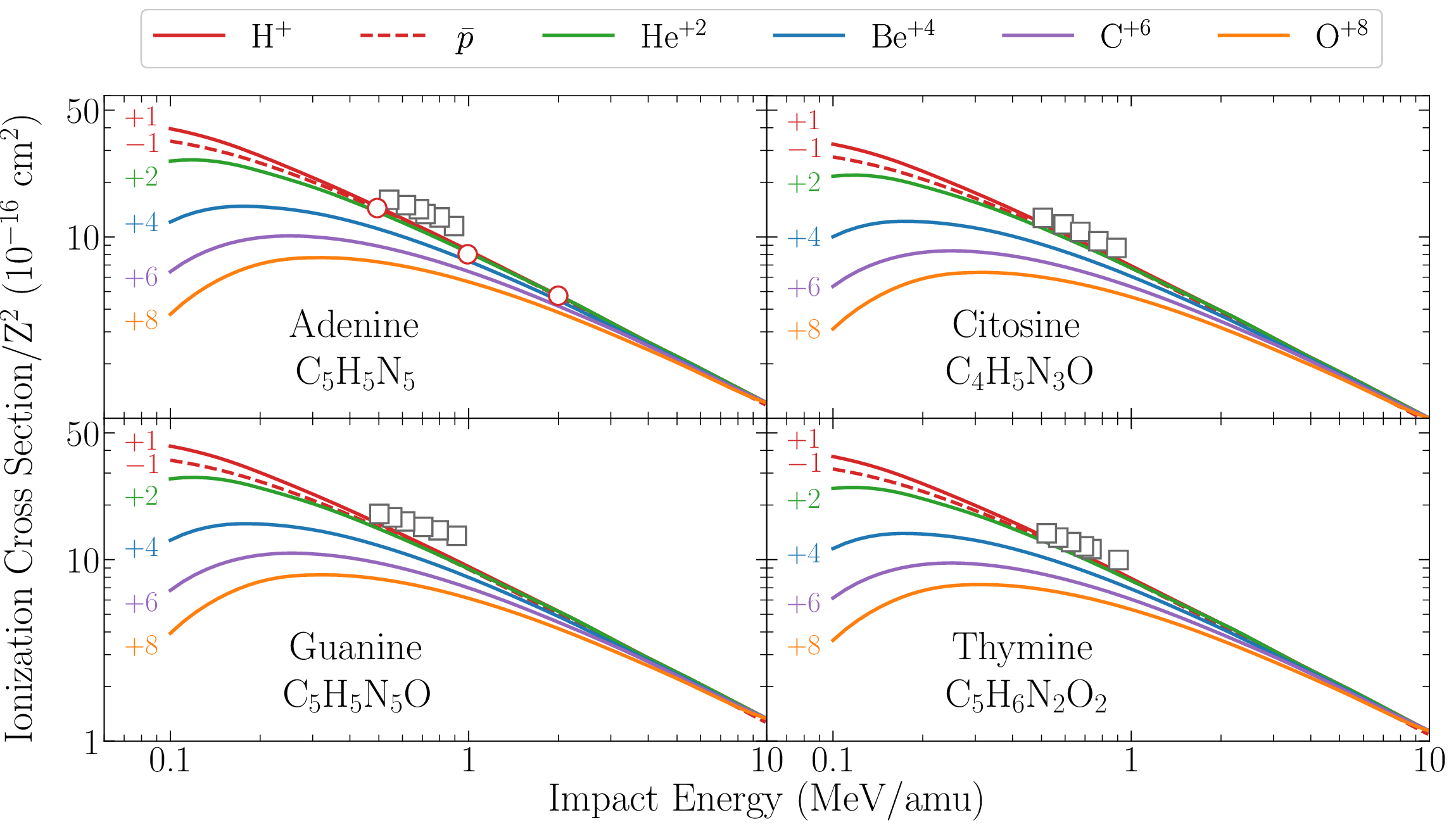}
\caption{Reduced CDW ionization cross section $\sigma_{M}/Z^2$ as a function 
of ion impact energy. Experiments: \mbox{\Large$\circ$}~\cite{iriki2011} 
for proton impact and $\square$~\cite{rahman2016} for electron impact 
with equivelocity conversion.}
\label{fig:crossDNA_1}
\end{figure*}

In Fig.~\ref{fig:crossDNA_1}, we report the reduced total ionization cross 
sections $\sigma_M/Z^2$ for adenine, cytosine, guanine, and thymine 
by the impact of multicharged ions obtained combining the SSM given by 
Eq.~(\ref{eq:sumion}) and the CDW results. For adenine, the agreement 
with the experimental 
data available for proton impact~\cite{iriki2011} is excellent. To the 
best of our knowledge, there are no experimental data on ion--collision 
ionization for the rest of the molecules. We have also included in this 
figure electron impact measurements~\cite{rahman2016} with the 
corresponding equivelocity conversion for electron incident energies 
higher than 300~eV. In this region, the proton and electron cross 
section should converge. Although the electron impact measurements are 
above our findings for all the molecular targets, it is worth stating 
that our results agree very well with other electron impact theoretical 
predictions~\cite{mozejko2003,tan2018}. 

\begin{figure*}[t!]
\centering
\includegraphics[width=0.8\textwidth]{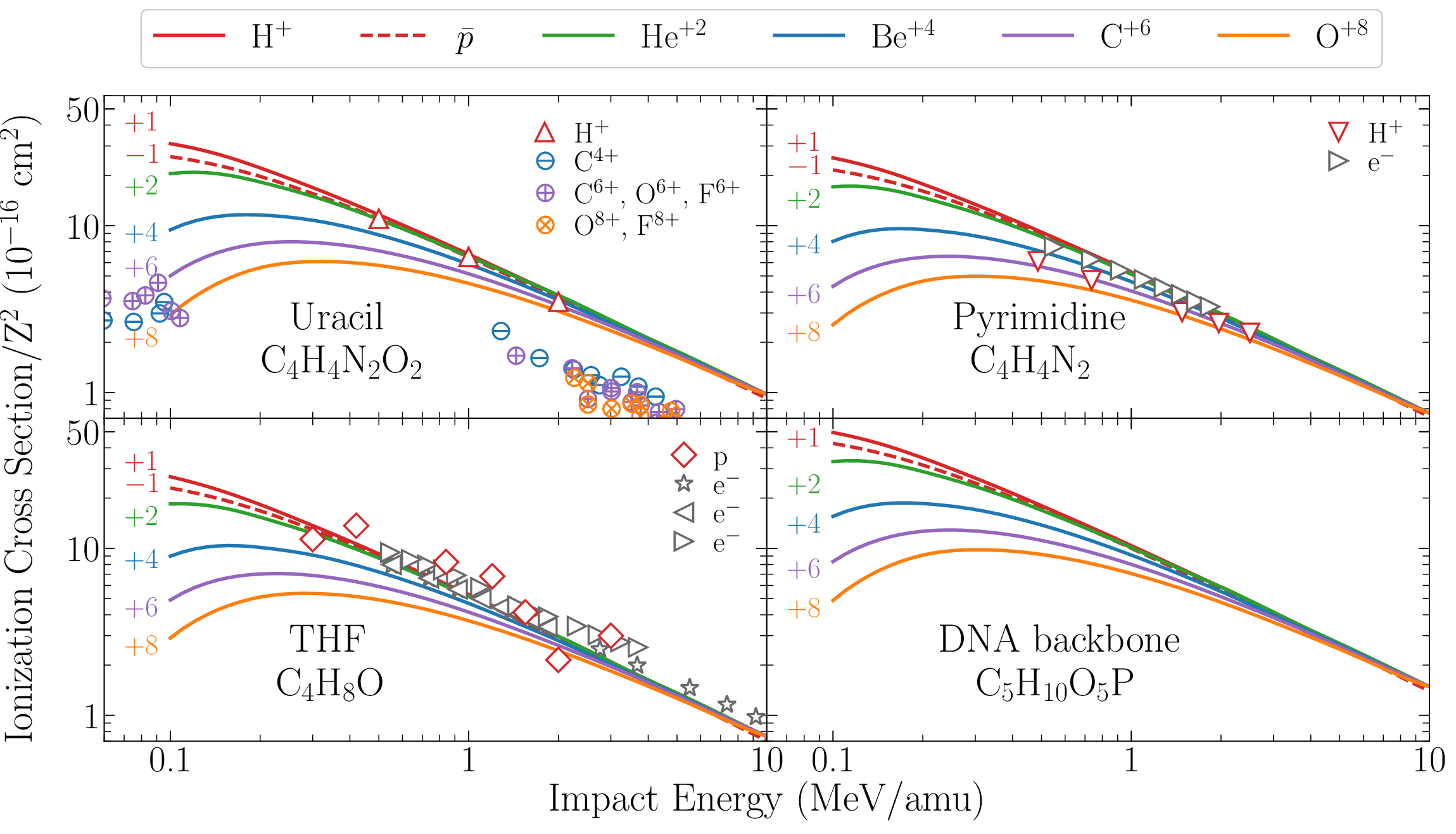}
\caption{Reduced CDW ionization cross section $\sigma_{M}/Z^2$ as a function 
of ion impact energy. Experiments: proton impact on 
$\triangle$ uracil~\cite{itoh2013}, 
$\bigtriangledown$ pyrimidine~\cite{wolff2014} and $\meddiamond$
THF~\cite{wang2016}. Impact of $\ominus$ C$^{4+}$, 
$\oplus$ C$^{6+}$, O$^{6+}$, F$^{6+}$, and
$\otimes$ O$^{8+}$, F$^{8+}$ on 
uracil~\cite{agnihotri2012,agnihotri2013}. 
Symbols~$\rhd$~\cite{bug2017}, $\lhd$~\cite{wolf2019}, and 
$\medstar$~\cite{fuss2009} for electron impact with equivelocity 
conversion.}
\label{fig:crossDNA_2}
\end{figure*}

The reduced total ionization cross sections $\sigma_M/Z^2$ for uracil, 
DNA backbone, pyrimidine, and THF are displayed 
in Fig.~\ref{fig:crossDNA_2}. For uracil, the agreement with the 
experimental proton impact measurements by 
Itoh~{\it et al.}~\cite{itoh2013} is good.
However, for the same target, our theory is a factor of two above 
the experimental ionization measurements by Tribedi and 
collaborators~\cite{agnihotri2012,agnihotri2013} by the impact of 
multicharged ions.
Nonetheless, it should be stated that our theoretical results coincide 
with calculations by Champion, Rivarola, and 
collaborators~\cite{agnihotri2012,champion2012}, which may indicate a 
possible misstep of the experiments. 

For pyrimidine, we show a comparison of our results with experimental 
data for proton impact by Wolff~\cite{wolff2014} and also for electron
impact ionization~\cite{bug2017} at high energies. 
The electron impact measurements 
agree with our calculations for energies higher than 500keV. 
Unexpectedly, the proton impact cross sections are significantly lower 
than our findings. 
Much more experiments are available for ionization of THF molecule by 
proton~\cite{wang2016} and by electron~\cite{bug2017,wolf2019,fuss2009} 
impact. Our SSM with CDW results show overall good agreement with these
data.

At intermediate impact energies, the $Z^2$ rule no longer holds, and 
other scalings can be considered in this region. For example, the 
molecular cross section and ion impact energy can be reduced with the
projectile charge $Z$, as suggested in in~\cite{janev1980,dubois2013}. 

\subsection{Scaling rules}
\label{subsec:scaling}
\subsubsection{Toburen rule}

The first attempt to develop a comprehensive but straightforward 
phenomenological model for electron ejection from large molecules was 
proposed by Toburen and coworkers~\cite{toburen1975,toburen1976}. 
The authors found it convenient to scale the experimental ionization 
cross section in terms of the number of weakly--bound electrons, $n_e$.
For instance, for C, N, O, P, and S, this number is the total number of 
electrons minus the K--shell. Following Toburen, the scaled ionization 
cross section per weakly bound electron $\sigma_{e}^T$ is
\begin{equation}
\sigma_{e}^T=\frac{\sigma_{M}}{n_e}\,, 
\label{27} 
\end{equation}
where $n_e=\sum_{\alpha}n_{\alpha}\nu_{\alpha}^T$, and $\nu_{\alpha}^T$ 
are the Toburen numbers given by
\begin{equation}
\nu_{\alpha}^T=\left\{ 
\begin{array}{ll}
1, & \text{for H,} \\
4, & \text{for C,} \\ 
5, & \text{for N and P,} \\ 
6, & \text{for O and S}\,.
\end{array}\right.
\label{eq:nelec} 
\end{equation} 
The Toburen rule can be stated by saying that 
$\sigma_{e}$ is a \textit{universal} parameter independent on the 
molecule, which depends solely on the impact velocity, and holds for 
high impact energies (i.e., 0.25--5 MeV/amu).
These $\nu_{\alpha}^T$ can be interpreted as the number of active 
electrons in the collision. At very high energies, the K--shell 
electrons will also be ionized, and these numbers will be different.
A similar dependence with the number of weakly bound electrons was 
found in Ref.~\cite{itoh2013} for proton impact on uracil and adenine.

\begin{figure*}[t!]
\centering
\includegraphics[width=0.8\textwidth]{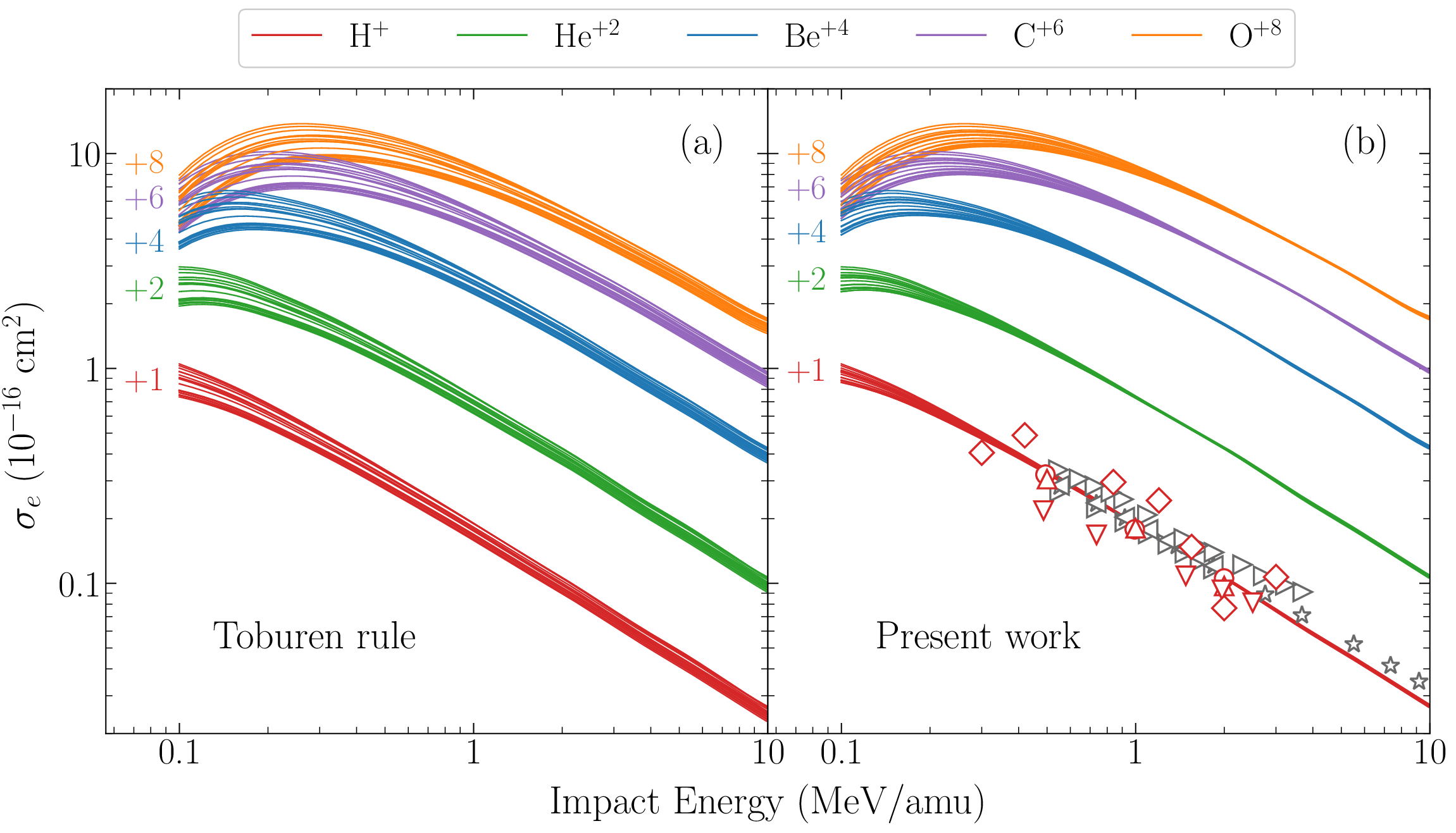}
\caption{Scaled ionization cross section per weakly bound electron using
(a)~the Toburen numbers $\nu_{\alpha}^T$, and (b) our proposed numbers
$\nu_{\alpha}^{\text{CDW}}$ for molecules listed in Table~\ref{tab:families}. 
For each band, the molecules are ordered from the smallest (top curve) to the
largest (bottom curve). Experiments: proton impact on 
\mbox{\Large$\circ$} adenine~\cite{iriki2011}, 
$\triangle$ uracil~\cite{itoh2013}, 
$\bigtriangledown$ pyrimidine~\cite{wolff2014} and $\meddiamond$ 
THF~\cite{wang2016}; electron impact on $\rhd$ pyrimidine~\cite{bug2017},
and $\lhd$, $\medstar$~\cite{wolf2019,fuss2009} THF.}
\label{fig:newscaling}
\end{figure*}

Following the Toburen scaling, we computed the scaled CDW cross sections 
$\sigma_{e}^T$ for the molecular targets of Table~\ref{tab:families}.
Our results are shown in Fig.~\ref{fig:newscaling}a as a function of 
the impact energy for different projectile charges. Although the 
Toburen scaling holds for high energies, its performance is still not 
satisfactory: the universal band is quite broad, as can be noted in 
this figure.

\subsubsection{CDW--based scaling}

The departure of our theoretical 
results from the Toburen rule can be easily understood 
by inspecting Fig.~\ref{fig:atomscaling}. It can be noted that the 
rule $\sigma_{\alpha}/\nu_{\alpha}^T\sim \sigma_{e}^T$, approximately 
constant, is not well satisfied by the CDW. 
For example, Fig.~\ref{fig:atomscaling} shows that the cross sections
for O are very similar to the cross sections for C, suggesting 
4 active electrons in O instead of 6. In the same way, the number of
active electrons for N, P, and S obtained with the CDW are also 
different from the $\nu_{\alpha}^T$ of Eq.~(\ref{eq:nelec}). 

Based on the CDW results, we propose a new scaling,
\begin{equation}
\sigma_{e}'=\frac{\sigma_M}{n_e'},
\label{32} 
\end{equation}
where $n_e'=\sum_{\alpha}n_{\alpha}\nu_{\alpha}^{\text{CDW}}$, and 
$\nu_{\alpha}^{\text{CDW}}$ are the numbers of active electrons
per atom obtained from the CDW ionization cross sections for 
different ions in H, C, N, O, P, and S targets, given as follows,
\begin{equation}
\nu_{\alpha }^{\text{CDW}} \sim\left\{ 
\begin{array}{ll}
1, & \text{for H,} \\
4, & \text{for C, N, and O,} \\ 
4.5, & \text{for P and S}\,.
\end{array}
\right. 
\label{eq:scalingCDW}
\end{equation}

The new scaled cross sections $\sigma_{e}'$ are plotted in 
Fig.~\ref{fig:newscaling}b. The experimental data for ionization of 
adenine~\cite{iriki2011}, uracil~\cite{itoh2013}, 
pyrimidine~\cite{wolff2014}, and THF~\cite{wang2016} by proton impact in
Fig.~\ref{fig:newscaling}b seems to corroborate the new scaling. 
We also included the electron impact ionization measurements with 
equivelocity conversion on pyrimidine~\cite{bug2017} and 
THF~\cite{bug2017,wolf2019,fuss2009}. 
It will be interesting to cross--check with future experiments, mainly 
for higher projectile charge states. 

\begin{table}[t]
\begin{center}
\begin{tabular}{|p{0.07\textwidth}p{0.02\textwidth}p{0.02\textwidth}|
p{0.095\textwidth}p{0.02\textwidth}p{0.02\textwidth}|
p{0.095\textwidth}p{0.035\textwidth}p{0.02\textwidth}|}
\hline
 Molecule & $n_e'$ & $n_e$ & Molecule          & $n_e'$ & $n_e$ & Molecule             & $n_e'$ & $n_e$ \\
\hline
 H$_2$    & 2 & 2   & C$_2$H$_7$N         & 19 & 20 & C$_4$H$_5$N$_3$O     & 37 & 42 \\
 H$_2$O   & 6 & 8   & C$_4$H$_8$O         & 28 & 30 & C$_5$H$_6$N$_2$O$_2$ & 42 & 48 \\
 NH$_3$   & 7 & 8   & C$_4$H$_4$N$_2$     & 28 & 30 & C$_5$H$_5$N$_5$      & 45 & 50 \\
 CH$_4$   & 8 & 8   & C$_6$H$_6$          & 30 & 30 & C$_5$H$_5$N$_5$O     & 49 & 56 \\
 CH$_5$N  & 13 & 14 & C$_4$H$_4$N$_2$O$_2$& 36 & 40 & C$_5$H$_{10}$O$_5$P  & 54.5 & 65 \\
 \hline
\end{tabular}
\caption{New scaling numbers $n_e'$, and Toburen numbers $n_e$,
for some molecular targets of biological interest.}
\label{nn}
\end{center}
\end{table}

\begin{figure*}[t!]
\centering
\includegraphics[width=0.785\textwidth]{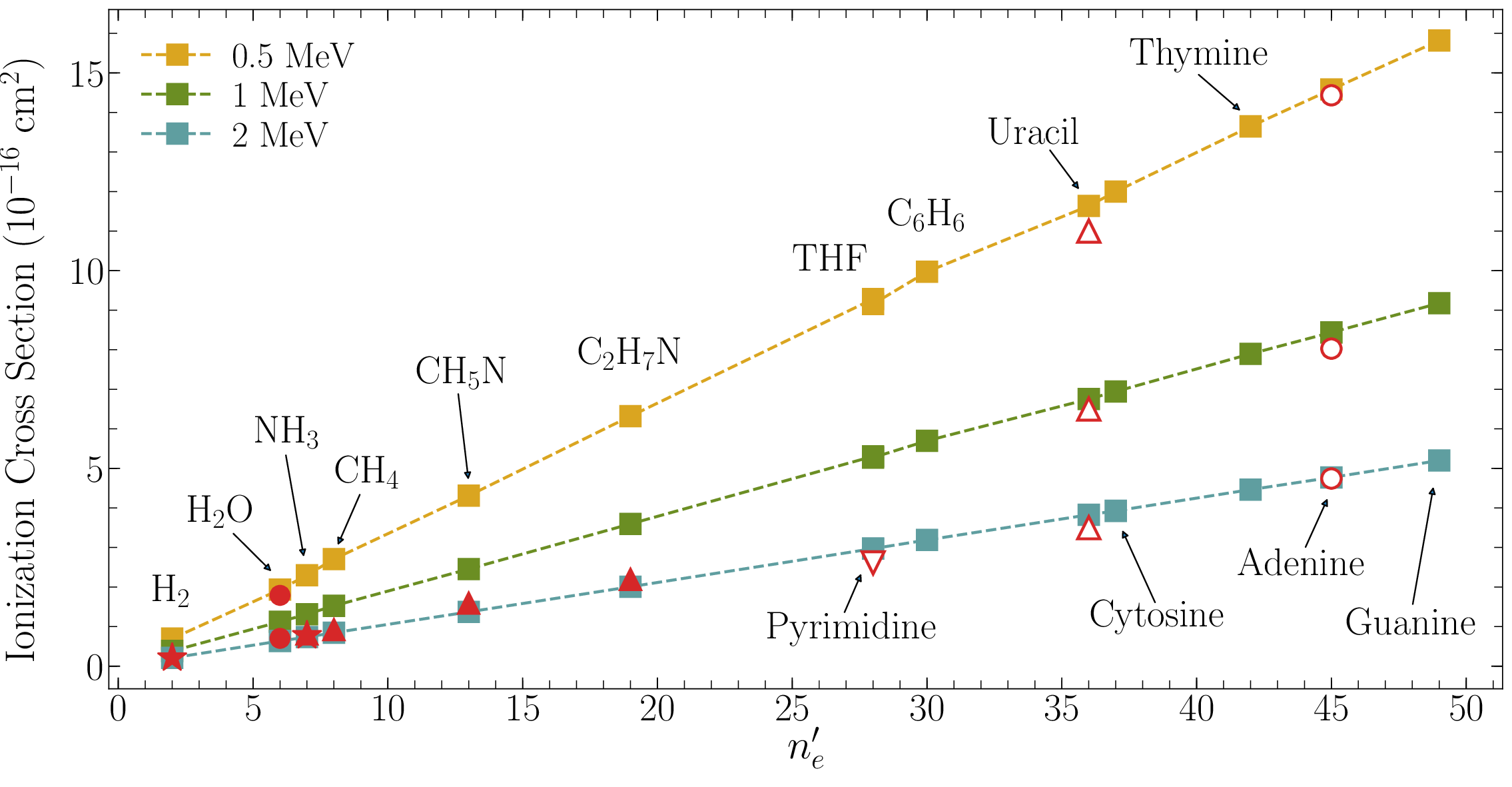}
\caption{Ionization cross sections by the impact of protons at 0.5, 1,
and 2 MeV in terms of the number of active electrons given by Table~\ref{nn}.
Experiments: 
\mbox{\Large$\circ$}~adenine~\cite{iriki2011}, 
$\triangle$ uracil~\cite{itoh2013}, 
$\bigtriangledown$ pyrimidine~\cite{wolff2014}, 
$\blacktriangle$ C$_2$H$_7$N, CH$_5$N, methane and ammonia~\cite{lynch1976},
\mbox{\scriptsize$\bigstar$} ammonia and H$_2$~\cite{rudd1985}, and 
\mbox{\Large$\bullet$} water~\cite{luna2007}.}
\label{fig:recta}
\end{figure*}

By using Eq.~(\ref{eq:scalingCDW}), we define new active electron 
numbers $n_e'$ for molecules. In Table~\ref{nn}, we display the present 
$n_e'$ values and $n_e$ ones by Toburen obtained from Eq.~(\ref{eq:nelec}).
Our values are different from the ones proposed by Toburen and used by 
other authors~\cite{itoh2013}, mainly due to the differences in the 
active electron numbers of oxygen. An alternative way of testing 
the present scaling can be attained by plotting the ionization cross sections 
of molecules as a function of the $n_e'$ from Table~\ref{nn}. Our 
findings are displayed in Fig.~\ref{fig:recta} for 
impact energies 0.5, 1, and 2 MeV. As can be noted, the computed CDW 
ionization cross sections for all the molecules show a linear 
dependence with the number of electrons $n_e'$ from Table~\ref{nn}.
We obtain similar results, even for $E=10$~MeV. The comparison with the 
experimental data available shows overall good agreement, for the 
smallest molecules, H$_2$, H$_2$O, and CH$_4$, up to the most complex 
ones, like adenine. For electron impact data, the experimental data was 
interpolated between close neighbors.
It is worth mentioning that an equivalent plot using the Toburen numbers 
$n_e$ does not exhibit the straight lines obtained with the present scaling.

While finishing the present work, we became aware of an accepted manuscript by 
L\"udde~{\it et al.}~\cite{ludde2019} on total ionization of biological
molecules by proton impact, using the independent--atom--model pixel counting
method~\cite{ludde2016,ludde2018}. The authors also raised a scaling 
with $\nu_{\alpha}=4$ 
for C, N, and O, but $\nu_{\alpha}=6$ for P. The agreement with this 
independent method for proton impact reinforces our multicharged--ion findings.

\subsection{Molecular structure of targets}
\label{subsec:molcalculations}

Finally, to test the range of validity of the SSM, we performed {\it ab initio}
molecular calculation of five nucleobases by employing the {\sc gamess} 
code. The geometry optimization and single point energy calculations 
were performed implementing the restricted Hartree--Fock method and the 
3-21G basis set. 

\begin{figure*}[t]
\centering
\includegraphics[width=0.8\textwidth]{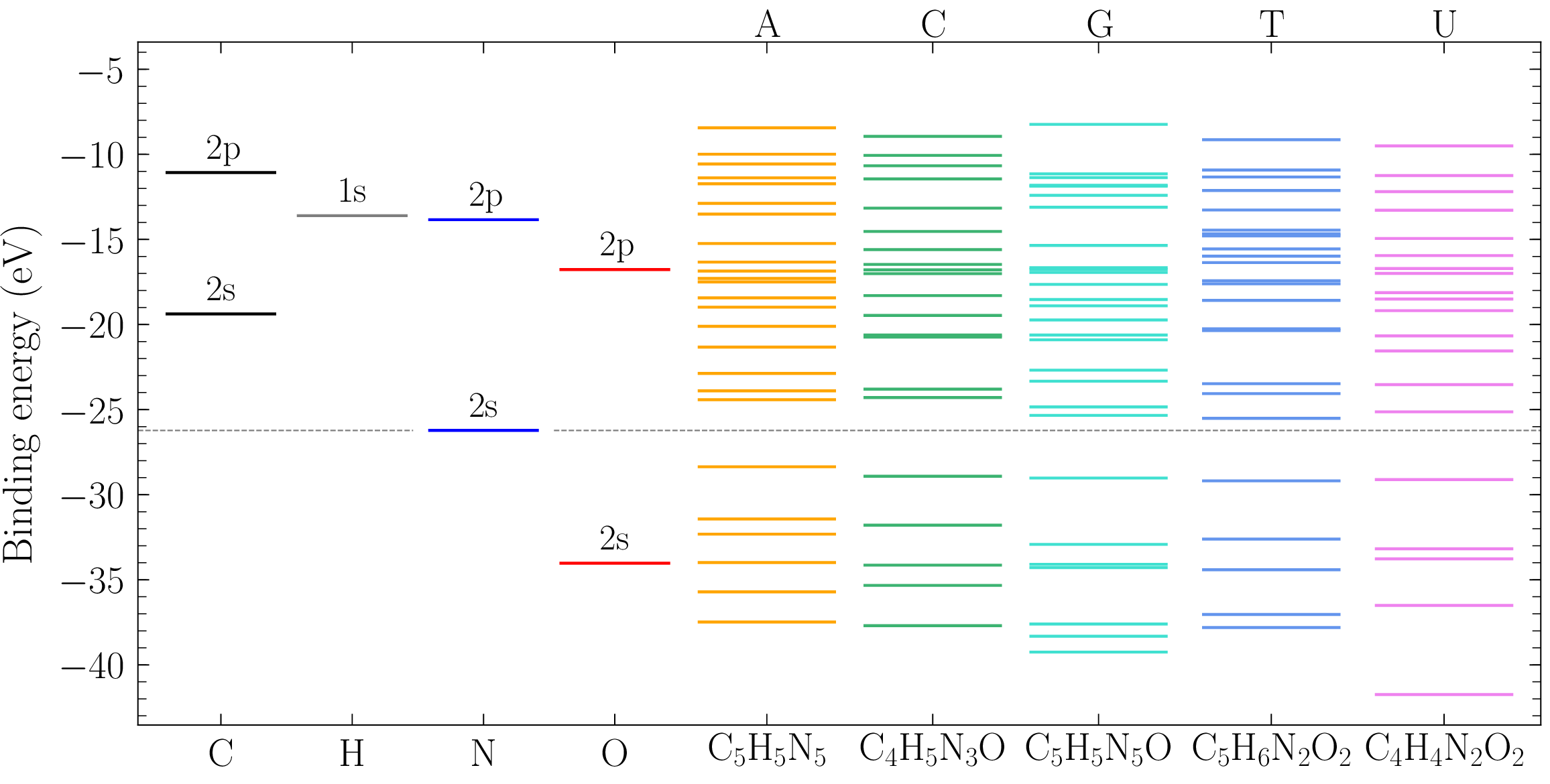}
\caption{Theoretical molecular binding energies for adenine, cytosine, 
guanine, thymine, and uracil compared to those of atomic constituents.}
\label{fig:bindener}
\end{figure*}

The molecular binding energies of the valence electrons for adenine, 
cytosine, guanine, thymine, and uracil are shown in Fig.~\ref{fig:bindener}. 
The binding energy of the highest molecular orbital (HOMO) agrees with 
the experimental values~\cite{Hush,Verkin,Dougherty} within 
2\% for all the DNA bases considered.
On the left side of Fig.~\ref{fig:bindener}, we show 
the atomic Hartree--Fock energies of the constituent elements, which 
gives an insight into the distribution of the weakly bound electrons
in the molecules. A dashed line around $-26$~eV is drawn to separate 
the molecular band in two. 
We can consider the atomic energy levels above this line as the ones
corresponding to the weakly bound electrons from Eq.~(\ref{eq:scalingCDW}).
For example, the $2s$ and $2p$ electrons of carbon are placed above
the separating line, which corresponds to the 4 electrons given by 
CDW--scaling. In the case of O, only the 4 electrons of 
the $2p$ orbitals are located above the separating line, which 
corresponds to the number of weakly bound electron given by 
our new scaling. 
The N case is not as straightforward; the $\nu_{N=4}^{\text{CDW}}$ 
would suggest that one out of the two $2s$ electrons contribute to the 
molecular scheme.

\subsubsection{A modified stoichiometric model}

The SSM considers the molecule to be assembled by isolated neutral atoms, 
which is definitively unrealistic. A first improvement can be suggested 
by assuming that the atoms are not neutral and that they have an uneven
distribution of electrons within the molecule, which can be expressed as 
an effective charge $q_{\alpha}$ per atom. The Mulliken charge gives a 
possible value for $q_{\alpha}$; however, there are a wide variety of 
charge distributions~\cite{lee2003}.

\begin{table*}[t]
\begin{center}
\begin{tabular}{|p{0.08\textwidth}|p{0.05\textwidth}|p{0.05\textwidth}|p{
0.05\textwidth}|p{0.05\textwidth}|p{0.19\textwidth}|}
\hline
Element & C & H & N & O & New stoichiometry \\
\hline
Adenine & +0.32 & +0.23 & --0.55 &       & 
C$_{4.92}$H$_{4.77}$N$_{5.14}$ \\ 
\hline
Cytosine & +0.28 & +0.21 & --0.56 & --0.53 & 
C$_{3.93}$H$_{4.79}$N$_{3.14}$O$_{1.13}$ \\ 
\hline
Guanine & +0.46 & +0.20 & --0.58 & --0.36 & 
C$_{4.89}$H$_{4.80}$N$_{5.15}$O$_{1.09}$ \\ 
\hline
Thymine & +0.20 & +0.19 & --0.54 & --0.52 & 
C$_{4.95}$H$_{5.81}$N$_{2.13}$O$_{2.13}$ \\ 
\hline
Uracil & +0.31 & +0.22 & --0.59 & --0.47 & 
C$_{3.92}$H$_{3.78}$N$_{2.15}$O$_{2.12}$ \\ 
\hline
\end{tabular}
\caption{Average effective Mulliken charge per atom $q_{\alpha}$, and 
new stoichiometric formula defined by Eq.~(\ref{eq:newstoi}) for five 
DNA molecules.}
\label{tab:newstoi}
\end{center}
\end{table*}

To take this effect into account, we can consider that the total amount 
of electrons $Q_{\alpha }$ on the element
$\alpha$ is equally distributed on all the $\alpha$ atoms. Therefore, 
each element $\alpha$ will have an additional charge, 
$q_{\alpha}=Q_{\alpha}/n_{\alpha}$, which can be positive or negative.
This amount will depend on the relative electronegativity respect to 
the other atoms~\cite{rappe1991}. 
Following this idea, we can estimate a new number of atoms per molecule
$n_{\alpha }^{\prime }$, given by
\begin{equation}
n_{\alpha }^{\prime }=n_{\alpha }-
\frac{q_{\alpha }}{\nu_{\alpha }^{\text{CDW}}}
\label{eq:newstoi}
\end{equation}%
In the case of neutral atoms, $q_{\alpha}=0$ and 
$n_{\alpha}^{\prime}=n_{\alpha}$, as it should be. 
In Table~\ref{tab:newstoi}, we display the average effective charge 
per atom $q_{\alpha}$ of C, H, N, and O, for five DNA molecules,
obtained from the full molecular calculation described 
above.

By implementing Eq.~(\ref{eq:newstoi}), it is possible to determine a 
new stoichiometric formula (last column of Table~\ref{tab:newstoi}). 
Now, instead of having an integer number of atoms $n_{\alpha}$, we obtain 
a fractional number $n_{\alpha}^{\prime}$. New molecular cross sections 
$\sigma^{\prime}_{M}=\sum_{\alpha}n_{\alpha}'\sigma_{\alpha}$ can be 
computed considering 
such values. Relative errors for the ionization cross sections were 
computed for the DNA bases from Table~\ref{tab:newstoi}. The differences 
obtained were less than 3\%, which indicates that the  
SSM is a quite robust model to handle these type of molecules within 
the range error expected for this model.

\section{Conclusions}

In this work, we have dealt with the calculation of ionization cross 
sections of seventeen biological molecules containing H, C, N, O, P, and 
S by the impact of antiprotons, H$^{+}$, He$^{2+}$, Be$^{4+}$, C$^{6+}$, 
and O$^{8+}$. To that end, we have employed the full CDW method and the 
simple stoichiometric model. 
The mean energy and angle of the emitted electrons, of importance in 
post--collisional radiation damage,  has also been calculated. Our 
findings show a clear dependence with the ion charge $Z$. For a given 
target as $Z$ increases, $\overline{E}_{\alpha}$ also increases, but 
$\overline{\theta}_{\alpha}$ decreases, showing a clear tendency to the 
forward direction. At impact energies greater than 2 MeV/amu, these 
values converge to the Born approximation, which embodies the simple 
$Z^{2}$ law. 

Total ionization cross sections for adenine, cytosine, thymine, guanine, 
uracil, DNA backbone, pyrimidine, and THF are presented and compared 
with the scarcely available experiments. We explored the rule of 
Toburen, which scales all the molecular ionization cross section 
normalizing with a certain number of weakly bound or valence electrons. 
We found that the ionization cross 
sections scales much better when normalizing with the number of active 
electrons in the collision obtained from the CDW results for atoms. 
This new scaling was tested with good results for the six 
projectiles and seventeen molecules studied here. The comparison with 
the experimental data reinforce our findings. Furthermore, we tested
the scaling by including experimental data of ionization of H$_2$, 
water, methane, and ammonia by proton impact showing good agreement at
intermediate to high energies.

Finally, we performed full molecular calculations for the DNA basis. 
By inspecting the molecular binding energy from quantum mechanical
structure calculations, we were able to understand the number of 
electrons proposed in our new CDW-based scaling. We attempt to improve 
the stoichiometric model by using the Mulliken charge to get fractional
rather than integer proportions. We found no substantial correction,
which indicates that the SSM works quite well.

In conclusion, the present results reinforce the 
reliability of the SSM to deal with complex molecules in the intermediate
to high energy range. Moreover, the simple stoichiometric model and the 
CDW cross sections in Ref.~\cite{miraglia2019} opens the possibility to 
describe a wide range of molecules containing H, C, N, O, P, and S, by 
the impact of multicharged ions.

\bigskip

\end{document}